\documentclass[aps,pra,twocolumn,superscriptaddress,longbibliography]{revtex4-1}

\usepackage{amsmath,amsfonts,amssymb,amsthm}
\usepackage{mathtools}
\usepackage{setspace}
\usepackage[normalem]{ulem} 
\usepackage{stmaryrd}       
\usepackage{dsfont}         
\usepackage{expdlist}
\usepackage{physics}
\usepackage{verbatim}
\usepackage{xspace}
\usepackage{tikz}
\usepackage{comment}

\usepackage{graphicx,xcolor}
\usepackage{hyperref}
\hypersetup{pdfstartview=}
\usepackage{url}

\usepackage{verbatim}
\usepackage{alltt}
\usepackage{moreverb}

\usepackage[ruled,lined]{algorithm2e}
\SetAlgorithmName{Box}{List of Protocols}

\providecommand{\ignore}[1]{}

\newif\ifcmnt
\cmnttrue
\ifdefined\cmntsoff\cmntfalse\fi

\ifcmnt
    \providecommand{\aucmnt}[1]{#1}

\else
    \providecommand{\aucmnt}[1]{}

\fi

\newcommand{\SI}[1]{\;\mathrm{#1}}

\newcommand{\Exp}{\mathbb{E}}

\newcommand{\TMPS}{\mathrm{TMPS}}

\newcommand{\epsx}{\epsilon_{x}}

\newcommand{\Rpow}[3]{\cR_{#1}\left(#2\middle|#3\right)}
\newcommand{\RpowB}[3]{\cR_{#1}\big(#2\big|#3\big)}

\newcommand{\Sfnt}[1]{\mathbf{#1}} 
\newcommand{\Pfnt}[1]{\mathsf{#1}} 

\newcommand{\cC}{\mathcal{C}}
\newcommand{\cD}{\mathcal{D}}
\newcommand{\cE}{\mathcal{E}}

\newcommand{\cM}{\mathcal{M}}

\newcommand{\cP}{\mathcal{P}}

\newcommand{\cR}{\mathcal{R}}

\newcommand{\cT}{\mathcal{T}}

\newcommand{\one}{\mathds{1}}

\newtheorem*{theorem*}{Theorem}

\newtheorem*{lemma*}{Lemma}

\allowdisplaybreaks

\begin{document}

\title{Experimental Low-Latency Device-Independent Quantum Randomness}

\author{Yanbao Zhang} \thanks{Y. Z. (yanbaoz@gmail.com) and L. K. S. contributed equally to this work.}
\affiliation{NTT Basic Research Laboratories and NTT Research Center for Theoretical Quantum Physics, NTT Corporation, 3-1 Morinosato-Wakamiya, Atsugi, Kanagawa 243-0198, Japan}
\author{Lynden K. Shalm} \thanks{Y. Z. (yanbaoz@gmail.com) and L. K. S. contributed equally to this work.}
\affiliation{National Institute of Standards and Technology, Boulder, Colorado 80305, USA}

\author{Joshua C. Bienfang}
\affiliation{National Institute of Standards and Technology, Gaithersburg, MD 20899, USA}
\author{Martin J. Stevens}
\affiliation{National Institute of Standards and Technology, Boulder, Colorado 80305, USA}
\author{Michael D. Mazurek}
\affiliation{National Institute of Standards and Technology, Boulder, Colorado 80305, USA}
\author{Sae Woo Nam}
\affiliation{National Institute of Standards and Technology, Boulder, Colorado 80305, USA}

\author{Carlos Abell\'{a}n}
\altaffiliation{Current address: Quside Technologies S.L., C/Esteve Terradas 1, Of. 217, 08860 Castelldefels (Barcelona), Spain}
\affiliation{ICFO-Institut de Ciencies Fotoniques, The Barcelona Institute of Science and Technology, 08860 Castelldefels (Barcelona), Spain}
\author{Waldimar Amaya}
\altaffiliation{Current address: Quside Technologies S.L., C/Esteve Terradas 1, Of. 217, 08860 Castelldefels (Barcelona), Spain}
\affiliation{ICFO-Institut de Ciencies Fotoniques, The Barcelona Institute of Science and Technology, 08860 Castelldefels (Barcelona), Spain}
\author{Morgan W. Mitchell}
\affiliation{ICFO-Institut de Ciencies Fotoniques, The Barcelona Institute of Science and Technology, 08860 Castelldefels (Barcelona), Spain}
\affiliation{ICREA-Instituci\'o Catalana de Recerca i Estudis Avan\c{c}ats, 08010 Barcelona, Spain}

\author{Honghao Fu}
\affiliation{Department of Computer Science, Institute for Advanced Computer Studies, and Joint Center for Quantum \break Information and Computer Science, University of Maryland, College Park, MD 20742, USA}
\author{Carl A. Miller}
\affiliation{Department of Computer Science, Institute for Advanced Computer Studies, and Joint Center for Quantum \break Information and Computer Science, University of Maryland, College Park, MD 20742, USA}
\affiliation{National Institute of Standards and Technology, Gaithersburg, MD 20899, USA}
\author{Alan Mink}
\affiliation{National Institute of Standards and Technology, Gaithersburg, MD 20899, USA}
\affiliation{Theiss Research, La Jolla, CA 92037, USA}
\author{Emanuel Knill}
\affiliation{National Institute of Standards and Technology, Boulder, Colorado 80305, USA}
\affiliation{Center for Theory of Quantum Matter, University of Colorado, Boulder, Colorado 80309, USA}

\begin{abstract}
  Applications of randomness such as private key generation and public randomness beacons require small blocks of certified random bits on demand.  Device-independent quantum random number generators can produce such random bits, but existing quantum-proof protocols and loophole-free implementations suffer from high latency, requiring many hours to produce any random bits.  We demonstrate device-independent quantum randomness generation from a loophole-free Bell test with a more efficient quantum-proof protocol, obtaining multiple blocks of $512$ random bits with an average experiment time of less than $5\SI{min}$ per block and with a certified error bounded by $2^{-64}\approx 5.42\times 10^{-20}$. 
\end{abstract}

\maketitle
\SetKwInOut{Calibration}{Calibration}
\SetKwInOut{RandAcc}{Randomness Accumulation}
\SetKwInOut{RandExt}{Randomness Extraction}
\SetKwInOut{Input}{Input}
\SetKwInOut{Output}{Output}
\SetInd{0.5em}{1em}

A fundamental feature of quantum mechanics is that measurements of a
quantum system can have random outcomes even when the system is in a
definite, pure state. By definition, pure states are completely
uncorrelated with every other physical system, which implies that the
measurement outcomes are intrinsically unpredictable by anyone outside
the measured quantum system's laboratory. The unpredictability of
quantum measurements is exploited by conventional quantum random number
generators (QRNGs)~\cite{herrero-collantes:qc2017a} for obtaining
random bits whose distribution is ideally uniform and independent of other systems.
The use of such QRNGs requires trust in the underlying quantum devices~\cite{pironio:qc2011a}.
A higher level of security is attained by device-independent quantum
random number generators (DIQRNGs)~\cite{colbeck:qc2006a,colbeck:qc2011c}
based on loophole-free Bell tests, where the randomness produced can be
certified even with untrusted quantum devices that may have been manufactured
by dishonest parties. The security of a DIQRNG relies on the physical
security of the laboratory to prevent unwanted information leakage,
and on the trust in the classical systems that record and process
the outputs of quantum devices for randomness generation.

Since the idea of DIQRNGs was introduced in Colbeck's
thesis~\cite{colbeck:qc2006a}, many DIQRNG protocols have
been developed---for a review see~\cite{acin:qc2016a}. These protocols
generally exploit quantum non-locality to certify entropy but differ
in device requirements, Bell-test configurations, randomness rates,
finite-data efficiencies, and the security levels achieved.
 We can classify protocols by whether they are secure in
the presence of classical or quantum side information, in other words,
by whether they are classical- or quantum-proof.

The first experimentally accessible DIQRNG protocol was given and implemented by
Pironio \emph{et al.}~\cite{pironio:qc2010a} with a detection-loophole-free
Bell test using entangled ions. They certified
$42$ bits of classical-proof entropy with error bounded by $0.01$, where,
informally, the error can be thought of as the probability
that the protocol output does not satisfy the certified claim.
This required about one month of experiment time.  To improve this result required
 the advent of loophole-free Bell tests and much more efficient
protocols. Such a protocol and experimental implementation with an
optical loophole-free Bell test was given by Bierhorst \emph{et al.}~\cite{bierhorst:qc2018a}
and obtained $1024$ classical-proof random bits with error $10^{-12}$ in $10\SI{min}$.
There have been three demonstrations of quantum-proof DIQRNGs,
all with photons.  The first two were subject to the locality
and freedom-of-choice loopholes~\cite{bell:qc1993a}.
 They obtained $4.6\times 10^{7}$ random bits with
error $10^{-5}$ in $111\SI{h}$~\cite{liu_yang:qc2017a}, and $6.2\times
10^{5}$ random bits with error $10^{-10}$ in
$43\SI{min}$~\cite{shen_lijiong:qc2018a}, respectively. The third was
loophole-free and obtained $6.2\times 10^{7}$ random bits with error
$10^{-5}$ in $96\SI{h}$~\cite{liu_yang:qc2018a}.

The quantum-proof experiments described above aimed for good asymptotic rates.
To approach the asymptotic rate requires a very large number of trials to certify
a large amount of entropy. However, many if not most applications of certified
randomness require only short blocks of fresh randomness.   To address these
applications, we consider instead a standardized request for $512$ random bits
with error $2^{-64}\approx 5.42\times 10^{-20}$ and with minimum delay,
or latency, between the request and delivery of  bits satisfying the request.
In this work, we consider only the contribution of experiment time to latency.
The previous quantum-proof DIQRNG implemented with a
loophole-free Bell test~\cite{liu_yang:qc2018a} would have required 
at least $24.1\SI{h}$ to satisfy the standardized
request---see Sect.~\ref{sect:eat} of the Supplemental Material (SM).

In this letter, we reduce the latency required to produce $512$ device-independent 
and quantum-proof random bits with error $2^{-64}$ by orders of magnitude. 
For this purpose, here we implement a quantum-proof protocol developed 
in the companion paper (CP)~\cite{knill:qc2018a} with a loophole-free Bell test.
Unlike other demonstrations of quantum-proof DIQRNGs,
we conservatively account for adversarial bias in
the setting choices, and we show repeated fulfillment of the
standardized request. We obtain five successive blocks of $512$ random
bits with error $2^{-64}$ and with an average experiment time
of less than $5\SI{min}$ per block.

\begin{figure}
\includegraphics[scale=.1]{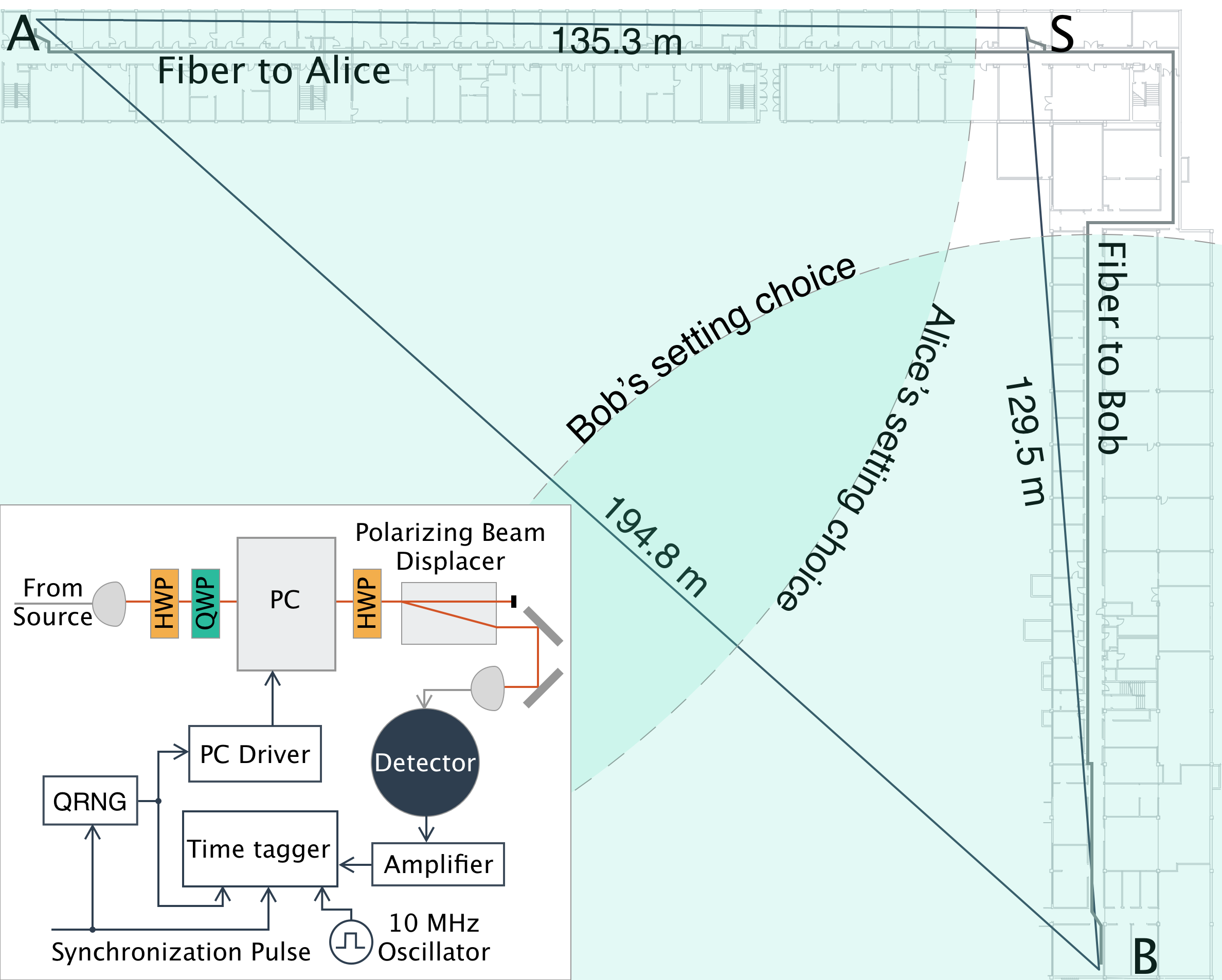}
\caption{Locations of Alice ($\Pfnt{A}$), Bob ($\Pfnt{B}$), and the source ($\Pfnt{S}$). Alice and Bob
 are separated by $194.8 \pm 1.0\SI{m}$ (this is slightly further than in 
 Refs.~\cite{shalm:qc2015a,bierhorst:qc2018a}). Faint grey lines indicate the 
 paths that the entangled photons take from the source to Alice and Bob through fiber optic cables.  
 The light-green quarter circles are the 2D projections of the expanding 
 light spheres containing the earliest available information about the random bits used for Alice's and Bob's setting 
 choices at the trial. When Bob finishes his measurement, the radius of the light sphere corresponding to the start of
  Alice's QRNG has expanded to $127.3 \pm 0.5\SI{m}$, after which it takes an additional $222.3 \pm 3.8\SI{ns}$ before the
  light sphere will intersect Bob's location. Similarly, when Alice completes her measurement, the light sphere 
  corresponding to the start of Bob's QRNG has only reached a radius of $98.3 \pm 0.5\SI{m}$, and it will take $315.5 \pm 
  3.8\SI{ns}$ more to arrive at Alice's station. In this way, the actions of Alice and Bob are spacelike separated. 
  Inset: Alice's and Bob's measurement apparatuses both consist of a Pockels cell (PC), operating at approximately 100 
  KHz, and a polarizer, constructed using two have-wave plates (HWPs), a quarter-wave plate (QWP) and a polarizing beam 
  displacer, in order to make fast polarization measurements on their respective photons. The measurement setting is 
  controlled by a QRNG,  the photon is detected by a high-efficiency superconducting
  nanowire single-photon detector, and the resulting signal is recorded on a time tagger, where a $10\SI{MHz}$ oscillator
  is used to keep Alice's and Bob's time taggers synchronized.}
\label{fig:layout}
\end{figure}

\noindent{\emph{Overview of theory.}}
We give a high-level description of the features of our protocol.
For formal definitions and technical details, see the CP~\cite{knill:qc2018a}.
Our protocol is based on repeated (but not necessarily independent or 
identical) trials of a loophole-free CHSH Bell test~\cite{clauser:qc1969a},
consisting of a source $\Pfnt{S}$ and two measurement stations $\Pfnt{A}$ and $\Pfnt{B}$
(see Fig.~\ref{fig:layout}).  In each trial, the source attempts to distribute
a pair of entangled photons to the stations, the protocol randomly chooses
binary measurement settings $X$ and $Y$ for the stations, the corresponding
measurements are performed, and the binary outcomes $A$ and $B$ are recorded.
We call $Z=XY$ and $C=AB$ the input and output of the trial, respectively.


An end-to-end randomness generation protocol starts with a request for $k$ random bits
with error $\epsilon$.  The user then chooses a positive quantity $\sigma$
(the entropy threshold for success) and positive errors $\epsilon_\sigma, \epsilon_x$
(the entropy error and the extractor error, respectively) whose sum is no more than $\epsilon$.
The quantity $\sigma$ chosen by the user must satisfy the inequality
$\sigma \geq  k+4\log_{2}(k)+4\log_{2}(2/\epsilon_{x}^{2})+6$.  
This inequality is sufficient to guarantee that, if the outputs of
the experiment can be proven to have entropy at least $\sigma$, then $k$  
random bits can be extracted. (The randomness extractor that we use for this purpose 
is Trevisan's extractor~\cite{trevisan:qc2001a} as implemented by 
Mauerer, Portmann and Scholz~\cite{mauerer:qc2012a}. We refer to it as the
TMPS extractor---see Sect.~\ref{sect:extractor} of the SM.)
The user also needs to decide the maximum number $n$ of Bell-test trials to run.
For simplicity, we temporarily assume that a fixed number $n$ of trials
will be executed, but in the implementation as described in a later
section we exploit the ability to stop early.

After fixing the parameters defined in the previous paragraph, $n$ Bell-test trials 
are sequentially executed, and the inputs and outputs are recorded
as $\Sfnt{Z}=(Z_{i})_{i=1}^{n}$ and $\Sfnt{C}=(C_{i})_{i=1}^{n}$, where $Z_i$
and $C_i$ are the input and output of the $i$'th trial. The upper-case symbols 
$\Sfnt{C}$, $C_i$, $\Sfnt{Z}$ and $Z_i$ are treated as random variables, and  
their values are denoted by the corresponding lower-case symbols. 
Let $\Pfnt{E}$ denote the ``environment'' of the experiment, including any quantum side information
that could be possessed by an adversary.  The entropy of the outputs $\Sfnt{C}$
is quantified by the quantum $\epsilon_{\sigma}$-smooth conditional min-entropy
of $\Sfnt{C}$ given $\Sfnt{Z}\Pfnt{E}$~\cite{renner:qc2006a}.  We
refer to this quantity as the output entropy. The user can estimate the
output entropy as described in the next section and check whether that estimate
is at least $\sigma$.  If not, the protocol fails and a binary variable $P$
is set to $P=0$; otherwise, the protocol succeeds and $P=1$.

When the protocol succeeds, we apply the TMPS extractor~\cite{mauerer:qc2012a}
to extract $k$ random bits with error $\epsilon$. The TMPS extractor is
a classical algorithm that is applied to the outputs $\Sfnt{C}$ as well as a
random seed $S$, and produces a bit string $R$. The final state of the protocol then
consists of the classical variables $RS\Sfnt{Z}P$ and the quantum system $\Pfnt{E}$.  
In the CP~\cite{knill:qc2018a}, we prove that the protocol is $\epsilon$-sound in the 
following sense: The error $\epsilon$ is an upper bound on the product of the success probability and the 
purified distance~\cite{tomamichel:qc2015a} between the actual state of $RS\Sfnt{Z}\Pfnt{E}$ 
conditional on the success event $P=1$ and an ideal state of $RS\Sfnt{Z}\Pfnt{E}$, 
according to which $RS$ is uniformly random and independent of $\Sfnt{Z}\Pfnt{E}$. 
For the protocol to be useful, it is necessary that the probability of success 
 in the actual implementation can be close to $1$, a property referred to as completeness.
With properly configured quantum devices, it is possible to make this probability exponentially
close to $1$ by increasing the number of trials executed. 
Soundness and completeness imply formal security of the protocol.

\noindent{\emph{Estimating entropy.}} In the CP~\cite{knill:qc2018a},  we develop
the approach of certifying entropy by ``quantum estimation factors'' (QEFs),
a general technique that generalizes previous certification techniques against
quantum side information~\cite{miller_c:qc2014b, arnon-friedman:qc2018a}. 
The construction of QEFs requires first defining a notion of models. 
The ``model'' for an experiment is  the set of all possible 
final states that can occur at the end of the experiment. A final state can be written as 
$\rho_{\Sfnt{CZ}\Pfnt{E}}=\sum_{\Sfnt{cz}}\ket{\Sfnt{cz}} \bra{\Sfnt{cz}}
\otimes \rho_{\Pfnt{E}}(\Sfnt{cz})$, where $\rho_{\Pfnt{E}}(\Sfnt{cz})$ 
is the unnormalized state of $\Pfnt{E}$ given results $\Sfnt{cz}$.   

Given the state $\rho_{\Sfnt{CZ}\Pfnt{E}}$, we characterize the unpredictability 
of the outputs $\Sfnt{c}$ given the system $\Pfnt{E}$ and the inputs $\Sfnt{z}$ by 
the sandwiched R\'enyi power, denoted by 
$\RpowB{1+\beta}{\rho_{\Pfnt{E}}(\Sfnt{cz})}{\rho_{\Pfnt{E}}(\Sfnt{z})}$ where
$\beta>0$ and $\rho_{\Pfnt{E}}(\Sfnt{z})=\sum_{\Sfnt{c}}\rho_{\Pfnt{E}}(\Sfnt{cz})$ 
(see Eq.~\eqref{eq:ren_power} of the SM for the explicit expression). A QEF with a 
positive power $\beta$ for a sequence of $n$ trials is a non-negative function $T$ 
of random variables $\Sfnt{CZ}$ such that for all states $\rho_{\Sfnt{CZ}\Pfnt{E}}$ 
in the model, $T$ satisfies the inequality
\begin{equation*}
  \sum_{\Sfnt{cz}} T(\Sfnt{cz}) \Rpow{1+\beta}{\rho_{\Pfnt{E}}(\Sfnt{cz})}{\rho_{\Pfnt{E}}(\Sfnt{z})}\leq 1.
\end{equation*}
Informally, one main result in the CP~\cite{knill:qc2018a} is that if at the conclusion
of the experiment the variable $\log_2 (T)/\beta$ takes a value at least $h$ for some $h > 0$, then the
output entropy (in bits) must be at least $ h - \log_2 ( 2 / \epsilon_\sigma^2 )/\beta$ no matter 
which particular state in the model describes the experiment. 
Hence, for estimating entropy it suffices to construct QEFs.

In practice, the model for a sequence of trials 
is constructed as a chain of models for each individual trial. QEFs then satisfy a chaining property:
If $F_i(C_iZ_i)$ is a QEF with power $\beta$ for the $i$'th trial, then the product
$\prod_{i=1}^n F_i ( C_iZ_i )$ is a QEF with power $\beta$ for the sequence of $n$ trials.
To construct the QEF $T(\Sfnt{CZ})$, we use this property. 
Moreover, since the model for each trial of our experiment is identical, we always take the 
same QEF for each executed trial.  The CP~\cite{knill:qc2018a} contains general
techniques for constructing models and QEFs, and the SM contains the details
of constructing models (Sect.~\ref{sect:thry}) and QEFs (Sect.~\ref{sect:calib}) for each 
trial of our experiment.

\begin{figure}
\includegraphics[scale=.1]{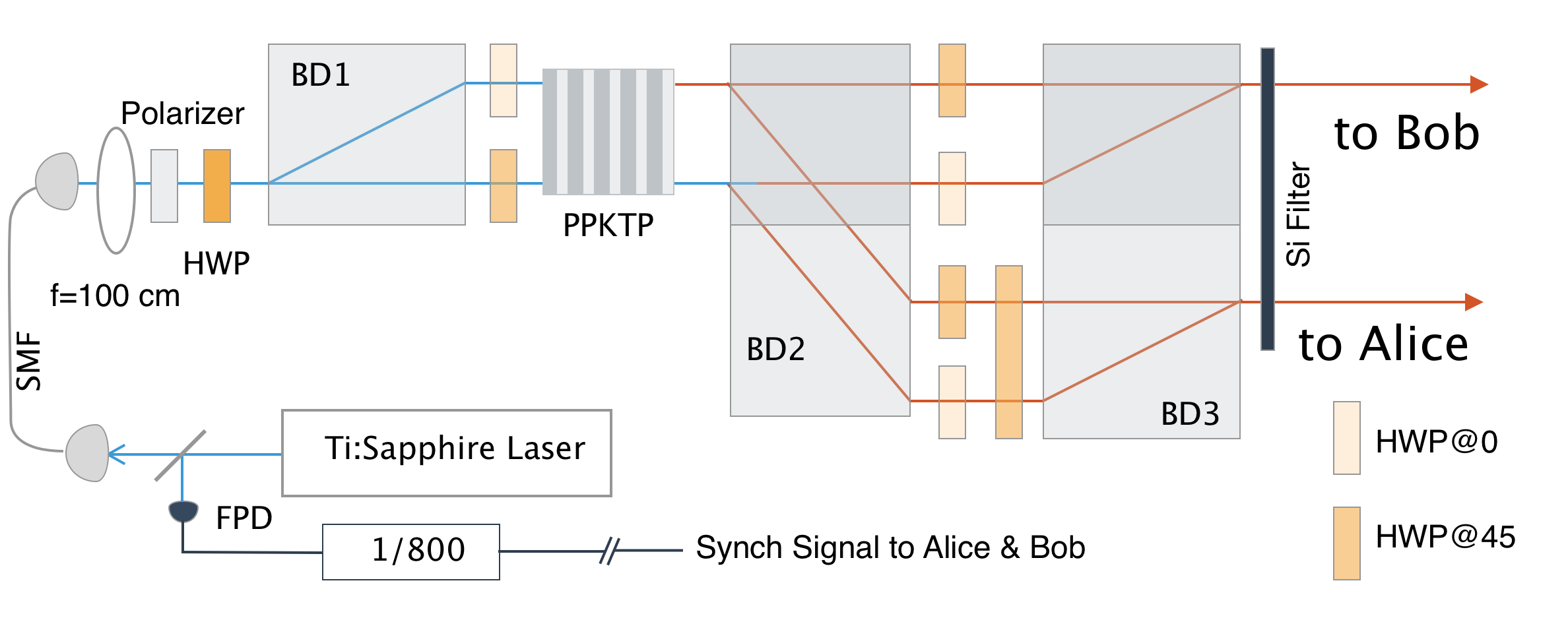}
\caption{Diagram of the entangled photon-pair source. A $775$-nm-wavelength picosecond Ti:Sapphire laser operating at
 a $79.3\SI{MHz}$  repetition rate pumps a $20$-mm-long periodically-poled potassium titanyl phosphate (PPKTP) crystal,
 to produce degenerate photons at $1550\SI{nm}$ with a per-pulse probability of $0.0045$. The pump is transmitted through 
 a polarization-maintaining single-mode fiber (SMF).  The PPKTP crystal is cut for type-II phasematching and placed in a 
 polarization-based Mach-Zehnder interferometer constructed using half-wave plates (HWPs) and three beam displacers (BD1,
 BD2 and BD3). Tuning the polarization of the pump by a polarizer and HWP allows us to create the non-maximally entangled 
 state $\ket{\psi}= 0.967 \ket{HH} +0.254 \ket{VV}$, where $H$ and $V$ denote the horizontally and vertically polarized 
 single-photon states. The photons, along with a synchronization signal, are then distributed via optical fiber to Alice  
 and Bob. The synchronization signal is generated by a fast photodiode (FPD) and divider circuit which divides the pump
 frequency by $800$, and is used as a clock to determine the start of a trial and to time the operation of Alice's and 
 Bob's measurements. This leads to a trial rate of approximately $100\SI{kHz}$.}
\label{fig:setup}
\end{figure}

\noindent{\emph{Experiment.}}  Our setup is similar to those reported in Refs.~\cite{shalm:qc2015a,bierhorst:qc2018a}. 
A pair of polarization-entangled photons are generated through the process of spontaneous parametric downconversion 
and then distributed via optical fiber to Alice and Bob (see Fig.~\ref{fig:setup}).  At each lab of Alice and Bob, 
a fast QRNG with parity-bit randomness extraction~\cite{abellan:qc2015a} is used to randomly switch a Pockels cell-based  polarization analyzer (see Fig.~\ref{fig:layout}). 
Alice's polarization measurement angles, relative to a vertical polarizer, are $a = −4.1^{\circ}$ and $a^{\prime} = 25.5^{\circ}$, and Bob's are $b = -a$ and $b^\prime = -a^\prime$. These measurement angles, along with the non-maximally entangled state prepared in Fig.~\ref{fig:setup}, are chosen based on numerical simulations of our setup to achieve an optimal Bell violation. The photons are then detected in each lab using superconducting nanowire single-photon detectors  with efficiency greater than $90\%$~\cite{Marsili2013}. The total system efficiencies for Alice and Bob are $76.2 \pm 0.3\%$ and $75.8 \pm 0.3 \%$, allowing the detection loophole to be closed. With the configuration detailed in Fig.~\ref{fig:layout}, 
we can also close the locality loophole. 

In each trial, Alice's and Bob's setting choices $X$ and $Y$ are made with random bits 
whose deviation from uniform is assumed to be bounded. That is,
knowing all events in the past light cone, one should not be able to
predict the next choice with a probability better than
$0.5+\epsilon_b$. We call $\epsilon_b$ the (maximum) adversarial bias.
 In particular, it is assumed that the quantum devices used
cannot have more prior knowledge of the random setting choices than
the adversarial bias for each trial. Specifically, we assume that the
adversarial and trial-dependent bias of Alice's and Bob's QRNGs is
bounded by $\epsilon_b \leq 1 \times 10^{-3}$. That is, each of the 
setting choices $X$ and $Y$ has a two-outcome distribution with 
probabilities in the interval $[0.5-1\times 10^{-3}, 0.5+1\times 10^{-3}]$. 
The bias assumption is supported in two ways: first by a quantum statistical model of the QRNGs,
validated by measurements of the QRNG internal operation~\cite{abellan:qc2015a},
and second by the observation that the frequencies of the output bits
of each QRNG deviate from 0.5 by less than $6 \times 10^{-5}$ on average 
in a run of $21\SI{min}$ of trials.

\noindent{\emph{Protocol implementation.}}  The goal is to obtain $k=512$ random
bits with error $\epsilon=2^{-64}$. For this, we set $\epsilon_{\sigma}=0.8\times 2^{-64}$ 
and $\epsilon_{x}=0.2\times 2^{-64}$. To extract $k=512$ random bits with the TMPS extractor, it suffices to
set the entropy threshold to be $\sigma=1089$. The implementation stages for each 
instance of the protocol are summarized in Box~\ref{prot:condgen_direct}, and more
details are available in Sect.~\ref{sect:imple} of the SM.

\RestyleAlgo{boxruled}
\vspace*{\baselineskip}
\begin{algorithm}
  \caption{Overview of protocol implementation}\label{prot:condgen_direct}
  \begin{enumerate}
    \item Calibration
          \begin{enumerate}
             \item Determine the QEF $F(CZ)$ and its power $\beta$ used for each executed trial.
             \item Fix $n$---the maximum number of trials.
          \end{enumerate} 
    
    \item Randomness Accumulation: Run the experiment to acquire up to $n$ trials. 
          After each trial $i$, 
                   \begin{enumerate}
                      \item Update the running $\log_{2}$-QEF value $L_{i}=\sum_{j=1}^{i}\log_{2}(F(c_{j}z_{j}))$, 
                            where $c_{j}$ and $z_{j}$ are the observed values of $C_{j}$ and $Z_{j}$.                             
                      \item If $\big(L_{i}-\log_{2}(2/\epsilon_{\sigma}^{2})\big)/\beta \geq \sigma$, 
                            stop the experiment, set the number of trials actually executed as $n_{\text{act}}=i$,   
                            and set the success event $P=1$. 
                   \end{enumerate}
    \item Randomness Extraction: If $P=1$, then extract $k$ random bits with error $\epsilon$.
  \end{enumerate}
\end{algorithm}

\noindent{\emph{Results.}}  Ideally, the protocol would be applied
concurrently with the acquisition of the experimental trials. In this
case, the trials were performed three months before the protocol was fully
implemented. About $89\SI{min}$ of experimental results were recorded.
The results were stored in $1\SI{min}$ blocks containing
approximately $6\times 10^{6}$ trials each.  The first $21\SI{min}$
were unblinded for testing the protocol, and the rest were kept
in blind storage until the protocol was fully implemented and ready
to be used.

From the first $21\SI{min}$ of unblinded results we decided to run 
five sequential instances of the protocol, and for calibration in each 
instance we determined to use the $10\SI{min}$  of results preceding 
to the first trial to be used for randomness accumulation (see Sect.~\ref{sect:imple} of the SM 
for details). We note that the trials for randomness accumulation in 
one instance can be used also for calibration in the next instance.   
For the protocol, we loaded the data and divided each $1\SI{min}$ block into $60$ 
subblocks of approximately $1\times 10^{5}$ trials each.  The protocol was then
designed to use integer multiples of these subblocks. 
The first instance of the protocol started producing randomness
at the $22$nd $1\SI{min}$ block. Each instance started at the first
not-yet-used subblock and used the previous $600$
subblocks for calibration, then processed subblocks until the running entropy
estimate surpassed the threshold $\sigma$. In each instance, this happened
well before the maximum number of trials $n$ determined at the calibration
stage was reached, leading to  success of the instance.
 We then applied the extractor to produce $512$ random bits with error $2^{-64}$.

\begin{table}
  \caption{Characteristics of the five protocol instances.
    The number of subblocks is approximately the number of seconds
    of experiment time required. 
    The entropy rate is estimated by $L_{n_{\mathrm{act}}}/(\beta n_{\mathrm{act}})$,
    where $n_{\mathrm{act}}$ is the actual number of trials
    executed in an instance, $L_{n_{\mathrm{act}}}$ is
    the running $\log_2$-QEF  value at the end of an instance,
    and $\beta$ is the power associated with the QEF which is used for 
    each executed trial and determined at the calibration stage. 
    The trial rate in the experiment was approximately $100\SI{kHz}$.}
  \label{tab:results}
  \begin{tabular}{|c|c|c|c|c|c|}
    \hline
      Instance   &  $n/10^{7}$ &
        $n_{\mathrm{act}}/10^{7}$ & Number& $\beta$ & Entropy \\
                &              &
                         & of sub-& & rate$/10^{-4}$ \\
                & &&blocks & &\\
    \hline
    1&  $5.25$ & $2.32 $ & $233$ & 0.010 & $6.07$ \\
    2&  $4.74$ & $3.76 $ & $379$ & 0.010 & $3.78$ \\
    3&  $5.92$ & $2.85 $ & $287$ & 0.009 & $5.47$ \\
    4&  $6.20$ & $2.83 $ & $285$ & 0.009 & $5.53$ \\
    5&  $5.49$ & $2.72 $ & $274$ & 0.010 & $5.20$ \\
    \hline
  \end{tabular}
\end{table}

The results are summarized in Tab.~\ref{tab:results}. It shows that
the experiment time required to fulfill the request for $512$ quantum-proof
random  bits with error $2^{-64}$ is less than $5\SI{min}$ on average,
demonstrating a dramatic improvement over other quantum-proof protocols and previous
experiments. The only experimentally accessible alternative quantum-proof protocol is
entropy accumulation as described in Ref.~\cite{arnon-friedman:qc2018a}. 
We found that satisfying the request using theoretical results from Ref.~\cite{arnon-friedman:qc2018a}, 
with our experimental configuration and performance, would have required at least $6.108\times 10^{10}$ 
trials, corresponding to $169.7 \SI{h}$ of experiment time---see Sect.~\ref{sect:eat} of the SM for details.  


In conclusion, we demonstrated five sequential instances of the DIQRNG protocol.  For
joint (or composable) security of the five instances,  it suffices that the quantum devices
do not retain memory of what happened during the previous instances. Without this assumption,
the joint security of the five instances can be compromised as explained in
Ref.~\cite{barrett_j:qc2012a}. In our implementation such problems are mitigated  
by the definition of soundness in terms of the purified distance rather than the
conventional trace distance, but the issues arising in composing 
protocols like ours need further investigation.

We have emphasized the importance of latency. To produce a fixed block
of random bits, latency is simply the time it takes for the protocol
to fulfill the request.    Above, we have neglected the
classical computing time required for calibration and extraction since
this can be made relatively small by using faster and more parallel
computers. For the current implementation the time costs for calibration 
and extraction are detailed in Sect.~\ref{sect:calib} and Sect.~\ref{sect:imple} of the SM, respectively.
The latency for our setup is limited by the rate at which we can implement
random setting choices, which in turn is limited by the Pockels
cells. Since the source produces pulses at a rate of $79.3\SI{MHz}$ and
 we can use $10$ successive laser pulses as a single trial without
reducing the quality of trials, if the Pockels cell limitation
can be overcome, the latency could be reduced by a factor of about $80$ with
the current entangled photon-pair source.

\begin{acknowledgments}
  This work includes contributions of the National Institute of
  Standards and Technology, which are not subject to U.S. copyright.
  The use of trade names does not imply endorsement by the
  U.S. government. The work is supported by the National
  Science Foundation RAISE-TAQS (Award 1839223);
  the European Research Council (ERC) projects AQUMET (280169), ERIDIAN (713682);
  European Union projects QUIC (Grant Agreement no.~641122) and FET Innovation Launchpad UVALITH (800901);
  the Spanish MINECO projects OCARINA (Grant Ref. PGC2018-097056-B-I00) and Q-CLOCKS (PCI2018-092973), the Severo Ochoa programme (SEV-2015-0522); Ag\`{e}ncia de Gesti\'{o} d'Ajuts Universitaris i de Recerca (AGAUR) project (2017-SGR-1354); Fundaci\'{o} Privada Cellex and Generalitat de Catalunya (CERCA program); Quantum Technology Flagship projects MACQSIMAL (820393) and QRANGE (820405); Marie Sk{\l}odowska-Curie ITN ZULF-NMR (766402); EMPIR project USOQS (17FUN03).
\end{acknowledgments}

%

%
%
%

\onecolumngrid
\newpage

\renewcommand{\theequation}{S\arabic{equation}}
\renewcommand{\thefigure}{S\arabic{figure}}


\section*{Supplemental Material: Experimental Low-Latency Device-Independent Quantum Randomness}
\label{sect:SM}

\section{Theory background}
\label{sect:thry}
We consider an experiment which has an input $Z$ and an output $C$ at each trial.
For the CHSH Bell-test configuration, the trial input consists of the random setting
choices $X$ and $Y$ of Alice and Bob, while the trial output consists of the corresponding
outcomes $A$ and $B$ of both parties. That is, $Z=XY$ and $C=AB$. The quantum state 
of the devices used in a trial is subsumed by the model below but does not appear explicitly. 
We therefore focus on the visible, classical variables $Z$ and $C$ referred to as
the trial results. The possible value that a classical variable takes is denoted by the corresponding
lower-case letter. There is an external quantum system $\Pfnt{E}$ carrying quantum side
information. We would like to certify randomness in $C$ with respect to $\Pfnt{E}$ and conditional on $Z$.
For this, we need to know the correlation between the trial results $CZ$ and the quantum system $\Pfnt{E}$.
After each trial of the experiment, the joint state of $CZ$ and $\Pfnt{E}$ is a
classical-quantum state
\begin{equation}\label{eq:cq_state}
\rho_{CZ\Pfnt{E}}=\sum_{cz}\ket{cz} \bra{cz} \otimes \rho_{\Pfnt{E}}(cz),
\end{equation}
where $\rho_{\Pfnt{E}}(cz)$ is the sub-normalized state of $\Pfnt{E}$ given trial results $cz$.
The trace $\tr\big(\rho_{\Pfnt{E}}(cz)\big)$ is the probability of observing the results
$cz$ at a trial.  In general, we consider the set of all possible classical-quantum states that can occur
at the end of the trial.  We refer to this set as the ``model'' $\cC$ for the trial. Similarly,
we can define the model for a sequence of trials. In this work, the phrase ``quantum state,'' unless 
otherwise specified, refers to a normalized quantum state.

We characterize the unpredictability of the output $c$ given the system $\Pfnt{E}$ and the input $z$ by the
sandwiched R\'enyi power, denoted by $\RpowB{1+\beta}{\rho_{\Pfnt{E}}(cz)}{\rho_{\Pfnt{E}}(z)}$, which is equal to 
\begin{equation}\label{eq:ren_power}
  \tr( (\rho_{\Pfnt{E}}(z)^{-\beta/(2+2\beta)}\rho_{\Pfnt{E}}(cz)\rho_{\Pfnt{E}}(z)^{-\beta/(2+2\beta)})^{1+\beta}),  
\end{equation} 
 where $\beta>0$ is a free parameter and $\rho_{\Pfnt{E}}(z)=\sum_{c}\rho_{\Pfnt{E}}(cz)$. 
 Our method relies on a class of non-negative functions $F: cz\mapsto F(cz)$, called ``quantum estimation factors'' 
 (QEFs). A QEF with power $\beta$ for a given trial is a non-negative function which satisfies the inequality
\begin{equation}\label{eq:qef_def}
  \sum_{cz} F(cz) \RpowB{1+\beta}{\rho_{\Pfnt{E}}(cz)}{\rho_{\Pfnt{E}}(z)} \leq 1
\end{equation}
at all states $\rho_{CZ\Pfnt{E}}$ in the trial model $\cC$.  Similarly, we can define a QEF with power $\beta$ 
for a sequence of trials given the model governing this sequence.
The above inequality is called the QEF inequality.

The concept of a QEF generalizes techniques for certifying randomness against quantum
side information used in previous works. The role of QEFs is similar to the role of 
the weighting terms in the weighted $(1+\epsilon)$-randomness function of Eq.~(6.4) in
Ref.~\cite{miller_c:qc2014b}, and also similar to the role of the quantum systems $D_i \overline{D}_i$ in
Eq.~(16) of Ref.~\cite{arnon-friedman:qc2018a}.  QEFs are also closely related to classical ``probability estimation factors'' (PEFs) as introduced in Refs.~\cite{knill:qc2017a,zhang:qc2018a}. 
When the quantum system $\Pfnt{E}$ has the minimum dimension of one, the sub-normalized states 
$\rho_{\Pfnt{E}}(cz)$ and $\rho_{\Pfnt{E}}(z)$ specify the probabilities $\mu_{\Pfnt{E}}(cz)$ and $\mu_{\Pfnt{E}}(z)$ 
of observing the results $cz$ and $z$ according to a distribution $\mu_{\Pfnt{E}}$. The model $\cC$ then captures
classical side information and specifies a set of probability distributions of $CZ$ given $\Pfnt{E}$.
In this case, the QEF inequality~\eqref{eq:qef_def} simplifies to
\begin{equation}\label{eq:pef_def}
  \sum_{cz} \mu_{\Pfnt{E}}(cz) F(cz) \mu_{\Pfnt{E}}(c|z)^{\beta}\leq 1.
\end{equation}
If a non-negative function $F: cz\mapsto F(cz)$ satisfies this inequality at all probability
distributions in the trial model $\cC$, then $F$ is a PEF with power $\beta$ 
for the trial~\cite{knill:qc2017a,zhang:qc2018a}.

The model $\cC$ for a trial is constructed as follows. Let $\Pfnt{D}$ be the quantum system of the devices 
used in the trial. The model $\cC$ is induced by a family of input-dependent positive-operator valued measures 
(POVMs) of $\Pfnt{D}$ with an input $Z$ that is 
``free'' in the sense that $Z$ is independent of other classical variables and the quantum systems $\Pfnt{D}, \Pfnt{E}$.
Before the trial, the joint state of the quantum systems $\Pfnt{D}$ and $\Pfnt{E}$ is described by a state
$\rho_{\Pfnt{DE}}$ which may depend on the previous trial results.  Let $\cP_{\Pfnt{D}, Z}(C)$ be a family of
$Z$-dependent POVMs of $\Pfnt{D}$ with outcome $C$. The specific family $\cP_{\Pfnt{D}, Z}(C)$ of POVMs
may depend on the previous trial results.  However, each POVM $P_{\Pfnt{D}, Z}(C)$ in $\cP_{\Pfnt{D}, Z}(C)$
should be consistent with the behavior of the quantum devices at the trial. In the CHSH Bell-test configuration, $Z=XY$,
$C=AB$, and the quantum system $\Pfnt{D}$ can be decomposed into two subsystems $\Pfnt{D}_1$ and $\Pfnt{D}_2$
held by Alice and Bob respectively. Hence, the POVM $P_{\Pfnt{D}, Z}(C)$ has a tensor-product structure over
the two subsystems $\Pfnt{D}_1$ and $\Pfnt{D}_2$. Furthermore, in a Bell test the non-signaling
conditions~\cite{PRBox, barrett:2005} are satisfied, so the output of a local party is independent
of the input of another local party. Therefore, for an arbitrary input $z=xy$ and output $c=ab$ the
POVM element is of the form $P_{\Pfnt{D}_{1},x}(a)\otimes P_{\Pfnt{D}_{2},y}(b)$
where $P_{\Pfnt{D}_{1},x}(A)$ and $P_{\Pfnt{D}_{1},y}(B)$ are POVMs. Given any input $z$,
the joint state $\rho_{C\Pfnt{E}|z}$ of the output $C$ and the system $\Pfnt{E}$ is induced by performing
a measurement $P_{\Pfnt{D}, z}(C)$ on the initial state $\rho_{\Pfnt{DE}}$. That is, for each $z$
\begin{equation} \label{eq:induced_state}
\rho_{C\Pfnt{E}|z}=\sum_c \ket{c} \bra{c}\otimes \tr_{\Pfnt{D}}\big(\rho_{\Pfnt{DE}} \left(P_{\Pfnt{D}, z}(c)\otimes \one_{\Pfnt{E}}\right) \big),
\end{equation}
where $\tr_{\Pfnt{D}}$ is the partial trace over the system $\Pfnt{D}$ and $\one_{\Pfnt{E}}$ is the
identity operator on the system $\Pfnt{E}$. The set of induced states $\rho_{C\Pfnt{E}|z}$ satisfying
the above physical constraints is denoted by $\cM(\cP_{\Pfnt{D}, z}(C); \Pfnt{E})$. Let $\cD(Z)$ be a set
of probability distributions of $Z$ at a trial. The specific set $\cD(Z)$ may depend on the previous
trial results.  If the input $Z$ is a free choice with distribution $\nu(Z)\in\cD(Z)$
and for each $z$ the state $\rho_{C\Pfnt{E}|z}$ is in $\cM(\cP_{\Pfnt{D}, z}(C); \Pfnt{E})$, then the final
state of the trial results $CZ$ and the quantum system $\Pfnt{E}$ is given by
\begin{equation} \label{eq:free_cq_state}
\rho_{CZ\Pfnt{E}}=\sum_{z}\nu(z)\ket{z} \bra{z}\otimes\rho_{C\Pfnt{E}|z}.
\end{equation}
 We construct the model $\cC$ governing each trial as the set of states of
 the above form with an appropriate set $\cD(Z)$ of input distributions 
 as specified in the following paragraph. We emphasize that although 
 a sequence of trials may be not independent and identically distributed (i.i.d.),
 the model governing each trial is the identical $\cC$. 

At each trial of our experiment, the input $Z=XY$, where $X$ and $Y$ are selected by QRNGs.
The distributions $\nu(X)$ and $\nu(Y)$ are each close to uniform. Specifically, they satisfy
$|\nu(x)-1/2|\leq \epsilon_{b}$ and $|\nu(y)-1/2|\leq \epsilon_{b}$ for all $x, y=0, 1$.
We call $\epsilon_{b}$ the (maximum) adversarial bias of the input random bits. For the model $\cC$,
we allow an arbitrary joint distribution $\nu(XY)$  as long as it lies in the convex envelope
of joint distributions of two independent binary variables where each variable's distribution satisfies
the above bias constraints.  It follows that the set $\cD(Z)$ of distributions of $Z=XY$ is a convex
polytope with $4$ extreme points. At these extreme points, the probability distributions are given by
$(p^2,pq, pq,q^2)$, $(pq, q^2,p^2,pq)$, $(pq, p^2,q^2,pq)$, and $(q^2,pq, pq,p^2)$ with $p=1/2+\epsilon_{b}$
and $q=1-p$, where a distribution $\nu(XY)$ is expressed as a vector $\big(\nu(X=0,Y=0), \nu(X=1,Y=0), \nu(X=0,Y=1), \nu(X=1,Y=1)\big)$. We denote these four extremal distributions by $\nu_k$, $k=1,2,3,4$. We note that the
convex polytope $\cD(Z)$ includes an open neighborhood of joint distributions at the uniform distribution, including
correlated ones. 
  

In view of the above construction of the model $\cC$, every state $\rho_{CZ\Pfnt{E}}\in\cC$ can be
written as a convex combination $\rho_{CZ\Pfnt{E}}=\sum_{k=1}^{4}\lambda_{k}\rho_{CZ\Pfnt{E}}^{(k)}$, 
where $\lambda_k\geq 0$,  $\sum_k \lambda_k=1$, and the states $\rho_{CZ\Pfnt{E}}^{(k)}$ can be expressed by
  Eq.~\eqref{eq:free_cq_state} with $\nu(z)$ replaced by $\nu_k(z)$. 
The model $\cC$ then admits a computationally accessible  characterization, see Thm.~5 of the 
companion paper (CP)~\cite{knill:qc2018a}.  
Based on this characterization, in Appendix G of the CP~\cite{knill:qc2018a} we presented an effective algorithm 
to compute a tight upper bound $f_{\max}$ on the sum $\sum_{cz}F'(cz)\RpowB{1+\beta}{\rho_{\Pfnt{E}}(cz)}{\rho_{\Pfnt{E}}(z)}$ for all states $\rho_{CZ\Pfnt{E}}$ in the model $\cC$ and for an arbitrary non-negative function 
$F': cz\mapsto F'(cz)$. From the definition of QEFs, one can see that the function $F:cz\mapsto F'(cz)/f_{\max}$ is 
a QEF with power $\beta$ for the model $\cC$. In this work, to construct a QEF with power $\beta$ we choose the 
non-negative function $F': cz\mapsto F'(cz)$ to be a PEF with the same power $\beta$, because not only are effective methods for constructing PEFs available but also PEFs exhibit unsurpassed finite-data efficiency~\cite{knill:qc2017a, zhang:qc2018a}. See Sect.~\ref{sect:calib} for details on the QEF construction. 

\section{Quantum-proof strong extractors}
\label{sect:extractor}

 Let $C$, $S$ and $R$ be classical variables with 
 the number of possible values denoted by $|C|$, $|S|$ and $|R|$, respectively.
 Define $m=\log_{2}(|C|)$, $d=\log_{2}(|S|)$ and $k=\log_{2}(|R|)$.
 When $C$, $S$ and $R$ are bit strings, $m$, $d$ and $k$ are their respective length.
 In the context of an extractor,  $C$ is its input, $R$ is its output, and $S$ is the seed, which
 has a uniform probability distribution and is independent of all other classical variables and
 quantum systems.  An extractor is specified by a function $\cE:(C,S)\mapsto R$.
 Before running the extractor, the joint state of $C$, $S$ and $\Pfnt{E}$
 is described as $\rho_{C\Pfnt{E}}\otimes \tau_{S}$, where
 $\rho_{C\Pfnt{E}} = \sum_{c} \ketbra{c}{c} \otimes \rho_{\Pfnt{E}}(c)$ and
 $\tau_{S}$ is a fully mixed state of dimension $2^{d}$.  After running the extractor,
 the joint state of $R$, $S$ and $\Pfnt{E}$ is described as
 $\rho_{RS\Pfnt{E}}=\sum_{rs} \ketbra{rs}{rs} \otimes \rho_{\Pfnt{E}}(rs)$.

 The function $\cE$ is called a quantum-proof strong extractor with parameters $(m,d,k,\sigma,\epsx)$
 if for every classical-quantum state $\rho_{C\Pfnt{E}}$ with quantum conditional min-entropy 
 $H_{\infty}(C|\Pfnt{E})\geq \sigma$ bits, the joint distribution of the extractor output $R=\cE(C,S)$
 and the seed $S$ is close to uniform and independent of $\Pfnt{E}$ in the sense that the 
 purified distance between $\rho_{RS\Pfnt{E}}$ and $\tau_{RS} \otimes \rho_{\Pfnt{E}}$ is less than or equal 
 to $\epsx$. Here $\tau_{RS}$ is a fully mixed state of dimension $2^{d+k}$ and $\rho_{\Pfnt{E}}$ is the marginal
 state of $\Pfnt{E}$ according to $\rho_{C\Pfnt{E}}$.

The above definition of quantum-proof strong extractors differs from others such as that in
Ref.~\cite{mauerer:qc2012a} by requiring small purified distance instead of small trace distance.
The definitions of both the purified and trace distances between two quantum states are
given in Sect.~3.2 of Ref.~\cite{tomamichel:qc2012a}. The purified distance can be extended
to the previously traced-out quantum systems such as that of the quantum devices used in the
protocol. This extendibility helps to analyze the composability of protocols involving the
same quantum devices, see Appendix A of the CP~\cite{knill:qc2018a} for detailed discussions.
We also note that as the purified distance is an upper bound of the trace distance (see
Prop.~3.3 of Ref.~\cite{tomamichel:qc2012a}), the above definition of quantum-proof
strong extractors implies the definition in Ref.~\cite{mauerer:qc2012a}.

To make the extractor work properly, the parameters $(m,d,k,\sigma,\epsx)$ need
to satisfy a set of constraints, called ``extractor constraints.'' The extractor constraints
always include that $1\leq \sigma\leq m$, $d\geq 0$, $k\leq \sigma$, and
$0<\epsx\leq 1$. A specific strong extractor with reasonably low seed requirements
is Trevisan's strong extractor~\cite{trevisan:qc2001a}, which is proved
to be quantum-proof in Ref.~\cite{de:2009}. Here we use Trevisan's strong extractor
based on the implementation of Mauerer, Portmann and Scholz~\cite{mauerer:qc2012a} that we refer to as the
TMPS extractor $\cE_{\TMPS}$. To run the TMPS extractor, additional extractor constraints are
\begin{align}\label{eq:tmps_con}
& k+4\log_2 (k) \le \sigma-6 +  4\log_2(\delta_x), \notag \\
& d \le w^2 \max \left(2, 1+
    \left\lceil\frac{\log_2(k-e)-\log_2(w-e)}{\log_2(e)-\log_2(e-1)}\right\rceil\right),
\end{align}
where $\delta_{x}$ is the desired upper bound on the trace distance between
$\rho_{RS\Pfnt{E}}$ and $\tau_{RS} \otimes \rho_{\Pfnt{E}}$, $w$ is the smallest
prime larger than $2\lceil\log_2(4mk^2/\delta_x^2)\rceil$, and $e$ is
 the base of the natural logarithm.  To ensure that the
purified distance is at most $\epsx$,  we set $\delta_x=\epsx^2/2$
according to the relation between the purified and trace distances as stated in
Prop.~3.3 of Ref.~\cite{tomamichel:qc2012a}. We remark that the first extractor
constraint in Eq.~\eqref{eq:tmps_con} is according to the $1$-bit extractor based on
polynomial hashing, which is directly from Ref.~\cite{mauerer:qc2012a}, while the second
extractor constraint is according to the block-weak design presented in Ref.~\cite{mauerer:qc2012a}
after considering the improved construction of a basic weak design of Ref.~\cite{ma:qc2012}.

\section{Details of protocol implementation}
\label{sect:imple}

Our goal is to obtain $k=512$ random bits with error $\epsilon=2^{-64}$. 
To achieve this goal, we set the smoothness error to be $\epsilon_{\sigma}=0.8\epsilon \approx 4.34\times 10^{-20}$
and the extractor error to be $\epsilon_{x}=0.2\epsilon \approx 1.08\times 10^{-20}$. We emphasize that
the positive errors $\epsilon_{\sigma}$ and $\epsilon_{x}$ need to satisfy that $\epsilon_{\sigma}+\epsilon_{x}\leq \epsilon$, but their choices are not unique. In order to reduce the number of trials 
(Eq.~\eqref{eq:exp_num_trials} of Sect.~\ref{sect:calib}) and the number of seed bits (Eq.~\eqref{eq:tmps_con} of 
Sect.~\ref{sect:extractor}) required to achieve the goal, we need to choose $\epsilon_{\sigma}$ and $\epsilon_{x}$ such that $\epsilon_{\sigma}+\epsilon_{x}=\epsilon$. Moreover, we observed that with the increase of the splitting ratio
$\epsilon_{\sigma}$:$\epsilon_{x}$, the number of trials required decreases while the number of seed bits
required increases. The splitting ratio $0.8$:$0.2$ used by us was not optimized; instead it was chosen heuristically 
such that it does not make the number of trials or the number of seed bits required too large. 
To satisfy the constraints of the TMPS extractor (see Eq.~\eqref{eq:tmps_con} of Sect.~\ref{sect:extractor}), the amount of quantum $\epsilon_{\sigma}$-smooth conditional min-entropy to be certified is $\sigma=1089$ bits. Below we describe
the stages required for implementing our protocol.

The first stage of the protocol is calibration based on the results preceding the
first trial to be used for randomness accumulation. 
To determine the number of trials required for a reliable calibration, we study the 
statistical strength, which is the minimum Kullback-Leibler divergence of 
the experimental distribution of trial results from the local realistic distributions
in a Bell test~\cite{vanDam:2005, zhang_y:qc2010a}. As explained in Ref.~\cite{zhang:qc2018a},
the latency for producing random bits is determined by the statistical strength:
the larger the statistical strength, the lower the latency becomes. 
From the first $21\SI{min}$ of unblinded results, we found that a stable estimate of 
the statistical strength needs at least $10\SI{min}$ of results. Consequently, a reliable 
calibration requires at least $10\SI{min}$ of results preceding the first trial to be used 
for randomness accumulation in each instance of the protocol.  As a result of the calibration 
stage, we determine a  well-performing  QEF $F(CZ)$ and its power 
$\beta$ used for each executed trial, and fix the maximum number of trials $n$ that 
can be used for randomness accumulation, see Sect.~\ref{sect:calib} for details. 

From the statistical strength determined from the first $21\SI{min}$ 
of unblinded results, we also estimated that an implementation of our protocol with 
a high probability of success requires about $8.75\SI{min}$ of trials with the trial 
rate $100\SI{kHZ}$ (see the values at the most left column of Tab.~\ref{tab:suc}). 
Considering that besides the first $21\SI{min}$ of unblinded trials we have about 
$68\SI{min}$ of trials left for implementing the protocol, 
we decided ahead of time to aim for five successful instances of the protocol. 

The second stage consists of acquiring up to $n$ trials.  After each
trial $i$, we update the running $\log_{2}$-QEF value
$L_{i}=\sum_{j=1}^{i}\log_{2}(F(c_{j}z_{j}))$, where $c_{j}$ and
$z_{j}$ are the actual values of variables $C_{j}$ and $Z_{j}$
observed at the $j$'th trial.  According to our theory, the output
  entropy estimated after the $i$'th trial is at least
  $\big(L_{i}-\log_{2}(2/\epsilon_{\sigma}^{2})\big)/\beta$. One advantage of
  QEFs~\cite{knill:qc2018a} is that we can stop the experiment early
  as soon as the running entropy estimate surpasses the threshold
  $\sigma$, that is, $\big(L_{i}-\log_{2}(2/\epsilon_{\sigma}^{2})\big)/\beta\geq \sigma$.
  If we fail to satisfy this condition after $n$ trials, the
protocol fails.  Let $n_{\mathrm{act}}$ be the actual number of trials
executed.

The third and final stage consists of applying the TMPS extractor to
the trial outputs. The extractor input is exactly $m=2n$ bits long and
consists of the trial outputs  padded with zeros to $2n$ bits if
$n_{\mathrm{act}} <n$. The amount of seed required by the
extractor is determined by $m$, $k$ and $\epsilon_{x}$ 
as instructed in Sect.~\ref{sect:extractor}. In each instance
of the protocol the number of seed bits provided to the extractor 
is $796322$, of which $398161$ bits were actually used. In our numerical 
implementation of the TMPS extractor, the extraction of $512$ random bits
with error $2^{-64}$ took about $3$ seconds on a personal computer for 
each  protocol instance.

\section{Calibration details}
\label{sect:calib}

Before each instance of the protocol we aim to minimize the number of trials required
to certify the desired amount of quantum smooth conditional min-entropy. For this,  
we first determine an input-conditional distribution $\nu(C|Z)$ by maximum likelihood using the 
calibration data (see Tab.~\ref{tab:calib}) and assuming i.i.d. calibration trials. 
We enforce the requirement that the distribution $\nu(C|Z)$ with $C=AB$ and $Z=XY$
satisfy non-signaling conditions~\citep{PRBox} and Tsirelson's bounds~\cite{Tsirelson:1980}. Denote
the set of conditional distributions satisfying non-signaling conditions and Tsirelson's bounds by $\cT_{C|Z}$,
and let the number of calibration trials with inputs $z=xy$ and outputs $c=ab$ be $n_{cz}$. 
Then, to obtain $\nu(C|Z)$ we need to solve the following optimization problem: 
\begin{equation}
  \begin{array}[b]{lll}
    \textrm{Max}_{\mu(C|Z)} & \sum_{cz} n_{cz}\log(\mu(c|z)) & \\
    \textrm{Subject to} & \mu(C|Z)\in \cT_{C|Z}. & 
  \end{array}
  \label{eq:maximum_likelihood}
\end{equation} 
The objective function is strictly concave and the set $\cT_{C|Z}$ is a convex polytope as characterized in Sect.~VIII 
of Ref.\cite{knill:qc2017a}, so there is a unique maximum, which can be found by convex programming. 
In our implementation we use sequential quadratic programming. The input-conditional distribution $\nu(C|Z)$ 
found for each protocol instance using the calibration data is shown in Tab.~\ref{tab:dists}. We remark that 
the above use of the i.i.d. assumption is only for determining the distribution $\nu(C|Z)$ in order to help 
the following QEF construction.

\begin{table}[htb!]
 \caption{Counts of measurement settings $xy$ and outcomes $ab$ used for calibration in the protocol.}\label{tab:calib}
 
 \begin{tabular}{c c}
 \\
  \begin{tabular}{|ll|l|l|l|l|}
    \hline
    \multicolumn{6}{|c|}{\text{Calibration data for Instance 1}} \\  
    \hline
    &$ab$&00&10&01&11\\
    $xy$&&&&&\\
    \hline
    00&&
    14828499 &   20247&   21081&   39893 \\
    10&&
    14700691 &   150422&   16012&   45361 \\
    01&&
    14685622 &   16396 &   165442&  44033 \\
    11&&
    14506915 &   191754 &   205253&   3425 \\
    \hline
  \end{tabular}

 &  
  \begin{tabular}{|ll|l|l|l|l|}
    \hline
    \multicolumn{6}{|c|}{\text{Calibration data for Instance 2}} \\  
    \hline
    &$ab$&00&10&01&11\\
    $xy$&&&&&\\
    \hline
    00&&
    14829111 &	20268 &	21486 &	40044 \\
    10&&
    14700512 &	150192 &	15731 &	45853\\
    01&&
    14685622 &	16371  &	164191 & 43981 \\
    11&&
    14510138 &	191978 &	203934 & 3452 \\
    \hline
  \end{tabular}
\\
\\

  \begin{tabular}{|ll|l|l|l|l|}
    \hline
    \multicolumn{6}{|c|}{\text{Calibration data for Instance 3}} \\  
    \hline
    &$ab$&00&10&01&11\\
    $xy$&&&&&\\
    \hline
    00&&
    14833584 &	20397 &	21730 &	39366 \\
    10&&
    14698516 &	149471 & 15704 & 45686 \\
    01&&
    14687682 &	16329  & 162921 & 43488 \\
    11&&
    14512332 &	191118 & 202908 & 3439 \\
    \hline
  \end{tabular}

&
  \begin{tabular}{|ll|l|l|l|l|}
    \hline
    \multicolumn{6}{|c|}{\text{Calibration data for Instance 4}} \\  
    \hline
    &$ab$&00&10&01&11\\
    $xy$&&&&&\\
    \hline
    00&&
    14831299 &	20421 &	21461 &	39383 \\
    10&&
    14694430 &	149505 & 15765 & 45042 \\
    01&&
    14677655 &	16275  & 163939 & 43348 \\
    11&&
    14505754 &	191564 & 204731 & 3432 \\
    \hline
  \end{tabular}
\\
\\

\multicolumn{2}{c}{
  \begin{tabular}{|ll|l|l|l|l|}
    \hline
    \multicolumn{6}{|c|}{\text{Calibration data for Instance 5}} \\  
    \hline
    &$ab$&00&10&01&11\\
    $xy$&&&&&\\
    \hline
    00&&
    14831005 &	20234 &	21422 &	39750 \\
    10&&
    14695631 &	149205 & 15729 & 44973 \\
    01&&
    14675545 &	16416  & 164758 & 43357 \\
    11&&
    14502760 &	192437 & 205327 & 3328 \\
    \hline
  \end{tabular}}
\end{tabular}
\end{table}

\begin{table}[htb!]
 \caption{The input-conditional distributions $\nu(C|Z)$ by maximum likelihood using the calibration data.
 They are used for determining trial-wise PEFs and QEFs, not to make a statement about the actual distribution 
 when running calibration or randomness accumulation in each instance of the protocol.}\label{tab:dists}
 
 \begin{tabular}{c}
 \\
  \begin{tabular}{|ll|l|l|l|l|}
    \hline
    \multicolumn{6}{|c|}{\text{The distribution $\nu(C|Z)$ for Instance 1}} \\  
    \hline
    &$ab$&00&10&01&11\\
    $xy$&&&&&\\
    \hline
    00&&
    0.994538669905741 &  0.001359201002169 &  0.001417406491026 &  0.002684722601064 \\
    10&&
    0.985821748235815 &  0.010076122672094 &  0.001071100768434 &  0.003031028323657 \\
    01&&
    0.984879607640748 &  0.001098577207454 &  0.011076468756019 &  0.002945346395779 \\
    11&&
    0.973101422088709 &  0.012876762759493 &  0.013791426915540 &  0.000230388236258 \\
    \hline
  \end{tabular}
  \\
  \\
 
  \begin{tabular}{|ll|l|l|l|l|}
    \hline
    \multicolumn{6}{|c|}{\text{The distribution $\nu(C|Z)$ for Instance 2}} \\  
    \hline
    &$ab$&00&10&01&11\\
    $xy$&&&&&\\
    \hline
    00&&
    0.994515705036610 &  0.001358847709653  & 0.001440882644962  & 0.002684564608775 \\
    10&&
    0.985817745162412 &  0.010056807583851  & 0.001054979429726  & 0.003070467824011 \\
    01&&
    0.984965456715605 &  0.001098349494673  & 0.010991130965968  & 0.002945062823755 \\
    11&&
    0.973168846092021 &  0.012894960118257  & 0.013703878500117  & 0.000232315289605 \\
    \hline
  \end{tabular}
  \\
  \\

  \begin{tabular}{|ll|l|l|l|l|}
    \hline
    \multicolumn{6}{|c|}{\text{The distribution $\nu(C|Z)$ for Instance 3}} \\  
    \hline
    &$ab$&00&10&01&11\\
    $xy$&&&&&\\
    \hline
    00&&
    0.994527969039707  & 0.001367319162871  & 0.001460067976166  & 0.002644643821256 \\
    10&&
    0.985882962094683  & 0.010012326107895  & 0.001051045129976  & 0.003053666667446 \\
    01&&
    0.985062951474202  & 0.001095311498405  & 0.010925085541671  & 0.002916651485723 \\
    11&&
    0.973323258322106  & 0.012835004650500  & 0.013610748902553  & 0.000230988124841 \\
    \hline
  \end{tabular}      
  \\
  \\

  \begin{tabular}{|ll|l|l|l|l|}
    \hline
    \multicolumn{6}{|c|}{\text{The distribution $\nu(C|Z)$ for Instance 4}} \\  
    \hline
    &$ab$&00&10&01&11\\
    $xy$&&&&&\\
    \hline
    00&&
    0.994550684213053 &  0.001368493402282  & 0.001440000126752  & 0.002640822257912 \\
    10&&
    0.985876463349061 &  0.010042714266275  & 0.001057058224524  & 0.003023764160141 \\
    01&&
    0.984968085635641 &  0.001092870990901  & 0.011022598704165  & 0.002916444669293 \\
    11&&
    0.973224019770634 &  0.012836936855908  & 0.013709501802950  & 0.000229541570507 \\
    \hline
  \end{tabular}      
  \\
  \\

  \begin{tabular}{|ll|l|l|l|l|}
    \hline
    \multicolumn{6}{|c|}{\text{The distribution $\nu(C|Z)$ for Instance 5}} \\  
    \hline
    &$ab$&00&10&01&11\\
    $xy$&&&&&\\
    \hline
    00&&
    0.994550644169521  & 0.001356542061317  & 0.001433498602964  & 0.002659315166198 \\
    10&&
    0.985858226009355  & 0.010048960221483  & 0.001057422898793  & 0.003035390870369 \\
    01&&
    0.984914018142560  & 0.001101986657382  & 0.011070124629925  & 0.002913870570133 \\
    11&&
    0.973153836105957  & 0.012862168693984  & 0.013761812802190  & 0.000222182397868 \\
    \hline
  \end{tabular}
  
\end{tabular}
\end{table}

Second, we determine the QEF and its power to be used at each executed trial for certifying randomness. For this, we assume that the quantum devices used are honest. Specifically, we assume that the trial results in the data to be analyzed are i.i.d. with the input-conditional distribution $\nu(C|Z)$ found above and with the uniform input distribution, that is, $p(z)=1/4$ for each $z=xy$.
We denote the distribution of each trial's results by $\nu(CZ)$, which is given as $\nu(C|Z)/4$. Given a QEF $F(CZ)$ with power $\beta$ and the target probability distribution $\nu(CZ)$ at each trial, according to our theory in the 
CP~\cite{knill:qc2018a} the amount of quantum $\epsilon_{\sigma}$-smooth  conditional min-entropy (in bits) available after $n$ trials in a successful implementation of our protocol is expected to be 
$n\Exp_{\nu}\log_{2}(F(CZ))/\beta-\log_2(2/\epsilon_{\sigma}^2)/\beta$, where $\Exp_{\nu}$ is the expectation functional  according to the distribution $\nu(CZ)$.  Therefore, the number of trials required to certify 
$\sigma=1089$ bits of quantum smooth conditional min-entropy with the smoothness error 
$\epsilon_{\sigma}=0.8\times 2^{-64}$ is given by
\begin{equation}\label{eq:exp_num_trials}
n_{\text{exp}}=\frac{\beta\sigma+\log_2(2/\epsilon_{\sigma}^2)}{\Exp_{\nu} \big(\log_2(F(CZ))\big)}.
\end{equation}
In principle, we can choose the QEF $F(CZ)$ and its power $\beta$ such that the number $n_{\text{exp}}$ is minimized. Such a QEF is optimal for our purpose. However, an effective algorithm for finding optimal QEFs has not yet been well developed. 
Instead, we determine a valid and well-performing QEF by a method described in the next paragraph.  

We replace the trial-wise QEF $F(CZ)$ with a trial-wise PEF $F'(CZ)$ with the same power $\beta$ in the above expression of $n_{\text{exp}}$, and we minimize $n_{\text{exp}}$ over the PEFs and the power $\beta$.  The PEF $F'(CZ)$ is constructed for the classical trial model  which includes all distributions of $CZ$ satisfying non-signaling conditions~\citep{PRBox}, Tsirelson's bounds~\cite{Tsirelson:1980},
and the specified adversarial bias $\epsilon_b$ with free setting choices. Denote the above classical trial model 
by $\cT_{CZ}$, which is a convex polytope as characterized in Sect.~VIII of Ref.\cite{knill:qc2017a}. Since the values 
of $\sigma$ and $\epsilon_{\sigma}$ are given, the minimization of $n_{\text{exp}}$ over the PEFs at a fixed $\beta>0$ is equivalent to the following maximization problem:
\begin{equation}
  \begin{array}[b]{lll}
    \textrm{Max}_{F'(CZ)} & \Exp_{\nu} \big(\log_2(F'(CZ))\big) & \\
    \textrm{Subject to} &  \sum_{cz} \mu(cz) F'(cz) \mu(c|z)^{\beta}\leq 1 \textrm{ for all } \mu(CZ)\in \cT_{CZ}, & \\
                        &  F'(cz)\geq 0  \textrm{ for all } cz. & 
  \end{array}
  \label{eq:optial_PEF}
\end{equation}
The objective function is strictly concave and the constraints are linear, so there is a unique maximum, which
can be found by the sequential qudratic programming (see Sect.~VIII of Ref.\cite{knill:qc2017a} for more details). 
After solving the minimization of $n_{\text{exp}}$ over the PEFs with a fixed $\beta>0$, the minimization over the 
power $\beta$ can be solved by any generic local search method. 
The optimal trial-wise PEF $F_s'(CZ)$ and its power $\beta_s$ found for each instance of our protocol are shown in 
Tab.~\ref{tab:pefs}. Once we obtain $F_s'(CZ)$ and $\beta_s$, 
according to the method discussed in Sect.~\ref{sect:thry} we can find the scaling factor $f_{\max}$
such that the function $F_s:cz\mapsto F_s'(cz)/f_{\max}$ is a valid QEF with power $\beta_s$ for each trial  
even considering the adversarial bias in the setting choices. We found that $f_{\max}$ is indistinguishable from $1$ 
at high precision. Specifically, we certified that $f_{\max}\in[1, 1+4\times 10^{-8}]$.  Thus, we can 
construct a well-performing trial-wise QEF in the sense that the constructed trial-wise QEF performs as well as 
the optimal trial-wise PEF used. 

\begin{table}[htb!]
 \caption{The optimal trial-wise PEF $F_s'(CZ)$ and its power $\beta_s$ constructed using the calibration data.}\label{tab:pefs}
 
 \begin{tabular}{c}
 \\
  \begin{tabular}{|ll|l|l|l|l|}
    \hline
    \multicolumn{6}{|c|}{\text{The PEF $F_s'(CZ)$ with $\beta_s=0.010$ for Instance 1}} \\  
    \hline
    &$ab$&00&10&01&11\\
    $xy$&&&&&\\
    \hline
    00&&
    0.999985100015945 &   0.960053330288753 &   0.961278860973820 &  1.031270546920231 \\
    10&&
    1.000014959703430 &   0.996179015633874 &   0.928539989152853 &  1.034730739709108 \\
    01&&
    1.000014959703431 &   0.929773555664518 &   0.996567940251360 &  1.036340302673597 \\
    11&&
    0.999984980337838 &   1.003805611257416 &   1.003418239233214 &  0.897122388776918 \\
    \hline
  \end{tabular}
  \\
  \\
 
  \begin{tabular}{|ll|l|l|l|l|}
    \hline
    \multicolumn{6}{|c|}{\text{The PEF $F_s'(CZ)$ with $\beta_s=0.010$ for Instance 2}} \\  
    \hline
    &$ab$&00&10&01&11\\
    $xy$&&&&&\\
    \hline
    00&&
    0.999983119719060 &  0.957736610299895 &  0.959750543337949  & 1.033066848043457 \\
    10&&
    1.000016947937376 &  0.995893262896469 &  0.924415087525900  & 1.035965231274906\\
    01&&
    1.000016947937377 &  0.926439911048989 &  0.996244181142510  & 1.038322042906155 \\
    11&&
    0.999982984135019 &  1.004090207359396 &  1.003740689984598  & 0.892082537083196 \\
    \hline
  \end{tabular}
  \\
  \\

  \begin{tabular}{|ll|l|l|l|l|}
    \hline
    \multicolumn{6}{|c|}{\text{The PEF $F_s'(CZ)$ with $\beta_s=0.009$ for Instance 3}} \\  
    \hline
    &$ab$&00&10&01&11\\
    $xy$&&&&&\\
    \hline
    00&&
    0.999987733298785 &   0.962390263422718 &  0.964371945028196 &  1.030101311154533 \\
    10&&
    1.000012315866351 &   0.996537925255370 &  0.932055378066285 &  1.032011535518689 \\
    01&&
    1.000012315866352 &   0.934047411232071 &  0.996837840568035 &  1.034285221532220 \\
    11&&
    0.999987634771458 &   1.003448155559641 &  1.003149437513694 &  0.903094436700780 \\
    \hline
  \end{tabular}      
  \\
  \\

  \begin{tabular}{|ll|l|l|l|l|}
    \hline
    \multicolumn{6}{|c|}{\text{The PEF $F_s'(CZ)$ with $\beta_s=0.009$ for Instance 4}} \\  
    \hline
    &$ab$&00&10&01&11\\
    $xy$&&&&&\\
    \hline
    00&&
    0.999988613440492 &  0.963372326842968  & 0.964857020693164  & 1.029377999040675 \\
    10&&
    1.000011432196966 &  0.996661005061451  & 0.933765419597876  & 1.031652352661214 \\
    01&&
    1.000011432197022 &  0.935258762949600  & 0.996997010088361  & 1.033467124616762 \\
    11&&
    0.999988521982605 &  1.003325574159495  & 1.002990910470073  & 0.905005556856297 \\
    \hline
  \end{tabular}      
  \\
  \\

  \begin{tabular}{|ll|l|l|l|l|}
    \hline
    \multicolumn{6}{|c|}{\text{The PEF $F_s'(CZ)$ with $\beta_s=0.010$ for Instance 5}} \\  
    \hline
    &$ab$&00&10&01&11\\
    $xy$&&&&&\\
    \hline
    00&&
    0.999986292840056  & 0.960621025921868 &  0.962460953372542 &  1.030949615008017 \\
    10&&
    1.000013762098517  & 0.996351569224136 &  0.929429804140644 &  1.033727107818069 \\
    01&&
    1.000013762098517  & 0.931280011007014 &  0.996713237820074 &  1.035921358844919 \\
    11&&
    0.999986182742716  & 1.003633756084664 &  1.003273531275535 &  0.898874175871150 \\
    \hline
  \end{tabular}
  
\end{tabular}
\end{table}

We emphasize that the above use of the i.i.d. assumption is only for determining a well-performing trial-wise QEF, 
while in our analysis of experimental data the i.i.d. assumption is not invoked.
To ensure that the probability of success in the actual implementation is high even if the experimental distribution of trial results $CZ$ drifts slowly with time, we conservatively set the maximum number of trials that can be used for randomness accumulation to $n=2n_{\textrm{exp},s}$, where $n_{\textrm{exp},s}$ is the number of trials required with the optimal PEF $F_s'(CZ)$ found in the above paragraph.  The values of $n$ at each instance are shown in Tab.~\ref{tab:suc}. If the quantum devices used are honest, we can bound the probability 
of failure at an instance with Bernstein's inequality~\cite{Bernstein}. The results are shown in Tab.~\ref{tab:suc}.
In the actual implementation of the protocol, each instance succeeded with an actual number of trials much less than $n$.
The data analyzed are presented in Tab.~\ref{tab:ana}.

\begin{table}[htb!]
 \caption{The maximum number, $n$, of trials required for each instance and the corresponding failure probability $p_{\text{fail}}$.}\label{tab:suc} 
 \begin{tabular}{c c c c c c}
 \hline
 Instance & 1 & 2 & 3 & 4 & 5 \\
 \hline 
 $n$     & 52481032 & 47374338 & 59237139 & 61990028 & 54890733 \\
 \hline
 $p_{\text{fail}}$ & $\leq 8.386\times 10^{-6}$ & $\leq 7.958\times 10^{-6}$ & $\leq 9.863\times 10^{-6}$ & $\leq 1.014\times 10^{-5}$ & $\leq 8.598\times 10^{-6}$ \\
 \hline
 \end{tabular}
\end{table}

\begin{table}[htb!]
 \caption{Counts of measurement settings $xy$ and outcomes $ab$ analyzed for randomness accumulation in the protocol.}\label{tab:ana}
 
 \begin{tabular}{c c}
 \\
  \begin{tabular}{|ll|l|l|l|l|}
    \hline
    \multicolumn{6}{|c|}{\text{Analysis data for Instance 1}} \\  
    \hline
    &$ab$&00&10&01&11\\
    $xy$&&&&&\\
    \hline
    00&&
    5766872 &	7890 &	8525 &	15483 \\
    10&&
    5715070 &	58133 &	6115 &	18096 \\
    01&&
    5713556 &	6361 &	62971 &	17067 \\
    11&&
    5643767 &	74691 &	78949 &	1334 \\
    \hline
  \end{tabular}

 &  
  \begin{tabular}{|ll|l|l|l|l|}
    \hline
    \multicolumn{6}{|c|}{\text{Analysis data for Instance 2}} \\  
    \hline
    &$ab$&00&10&01&11\\
    $xy$&&&&&\\
    \hline
    00&&
    9365500 &	12916 &	13661 &	24706 \\
    10&&
    9278437 &	94378 &	9907 &	28542\\
    01&&
    9269918 &	10273 &	103158 & 27282 \\
    11&&
    9160334 &	120357 & 128237 & 2185 \\
    \hline
  \end{tabular}
\\
\\

  \begin{tabular}{|ll|l|l|l|l|}
    \hline
    \multicolumn{6}{|c|}{\text{Analysis data for Instance 3}} \\  
    \hline
    &$ab$&00&10&01&11\\
    $xy$&&&&&\\
    \hline
    00&&
    7098856 &	9769 &	10200 &	19035 \\
    10&&
    7033040 &	71534 &	7528 &	21465 \\
    01&&
    7025429 &	7822 &	78637 &	20731 \\
    11&&
    6941352 &	92273 &	98527 &	1607 \\
    \hline
  \end{tabular}

&
  \begin{tabular}{|ll|l|l|l|l|}
    \hline
    \multicolumn{6}{|c|}{\text{Analysis data for Instance 4}} \\  
    \hline
    &$ab$&00&10&01&11\\
    $xy$&&&&&\\
    \hline
    00&&
    7044516 &	9510 &	10216 &	18839 \\
    10&&
    6981677 &	70746 &	7461 &	21440 \\
    01&&
    6969396 &	7845 &	78520 &	20625 \\
    11&&
    6889053 &	91212 &	97340 &	1582 \\
    \hline
  \end{tabular}
\\
\\

\multicolumn{2}{c}{
  \begin{tabular}{|ll|l|l|l|l|}
    \hline
    \multicolumn{6}{|c|}{\text{Analysis data for Instance 5}} \\  
    \hline
    &$ab$&00&10&01&11\\
    $xy$&&&&&\\
    \hline
    00&&
    6768897 &	9374 &	9996 &	18188 \\
    10&&
    6708625 &	68397 &	7033 &	20723 \\
    01&&
    6702989 &	7421 &	74355 &	19950 \\
    11&&
    6622018 &	87747 &	92572 &	1602 \\
    \hline
  \end{tabular}}

\end{tabular}
\end{table}

In our numerical implementation, the time cost for finding the maximally likely input-conditional
distribution $\nu(C|Z)$ and the optimal PEF $F_s'(CZ)$ with its power $\beta_s$ at each instance
of the protocol was about two seconds on a personal computer, which is negligible.
However, it took time to determine tight bounds on $f_{\max}$ in order to ensure
that the performance of the resulted QEF is as close as possible to that of the PEF used. We 
recall that as the same QEF is used for each executed trial, we need only to perform the certification 
of $f_{\max}$ once at each instance of the protocol. For this, we  
implemented the algorithm presented in Appendix G of the CP~\cite{knill:qc2018a} with parallel
computation in Matlab. According to the algorithm, the least upper bound and the greatest
lower bound on $f_{\max}$ are iteratively updated. At each iteration, we first need to divide
a 2-dimensional searching region into $t$ subregions and perform a computation
for each subregion independently. Then the bounds on $f_{\max}$ could be updated according
to the algorithm. This division and computation step can be implemented in parallel.
The parameter $t$ is free and reflects the tradeoff between the time cost and the
computational resource cost. In our implementation, we used $81$ parallel workers and so
we set $t=81$.  At each instance of the protocol, the certification that
$f_{\max}\in[1, 1+4\times 10^{-8}]$ at the numerical precision of $2^{-52}\approx 2.22\times 10^{-16}$
with Matlab took about $39\SI{min}$. We also verified the obtained bounds on $f_{\max}$ with Mathematica 
at the precision of $10^{-32}$. This verification consumed about $4.5\SI{min}$ on a personal computer for
each instance.

\section{Performance of entropy accumulation with CHSH-based min-tradeoff functions}
\label{sect:eat}

The entropy accumulation protocol as described in Ref.~\cite{arnon-friedman:qc2018a} is another experimentally 
accessible protocol for certifying smooth conditional min-entropy with respect to quantum side information.
The implementation of entropy accumulation requires a ``min-tradeoff function'' $f_{\min}$.
We studied the performance of entropy accumulation with the class of min-tradeoff functions
in Ref.~\cite{arnon-friedman:qc2018a}. These min-tradeoff functions are constructed 
from a lower bound on the single-trial conditional von Neumann entropy derived in 
Refs.~\cite{acin:2007, pironio:2009}. The lower bound is characterized as a function 
of the violation of the CHSH Bell inequality~\cite{clauser:qc1969a} (hence we are calling them
``CHSH-based min-tradeoff functions'').    
Given the expected violation $(\hat{I}-2)>0$ of the CHSH Bell inequality, a lower bound $\kappa$ on the
success probability of the entropy accumulation protocol, and the smoothness error $\epsilon_{\sigma}$,
the minimum number of i.i.d. trials, where the input distribution is uniform, required to
certify $\sigma$ bits of quantum smooth conditional min-entropy according to 
entropy accumulation with CHSH-based min-tradeoff functions is denoted by $n_{\textrm{EAT},\sigma}$.
The explicit expression for $n_{\textrm{EAT},\sigma}$ is given in Eq.~(S34) of our previous
work~\cite{zhang:qc2018a}, which is derived from the results presented in Ref.~\cite{arnon-friedman:qc2018a}. 
For convenience and completeness, we restate the result as follows:
\begin{equation} \label{eq:opt_EAT_trial_bound}
n_{\mathrm{EAT},\sigma}=\min_{3/4\leq p_t\leq (2+\sqrt{2})/4}n_{\mathrm{EAT},\sigma}(p_t),
\end{equation}
where $n_{\mathrm{EAT},\sigma}(p_t)$ is defined by
\begin{align} \label{eq:eat_rate}
&g(p) =  \begin{cases}
			 1 - h\left( \frac{1}{2} + \frac{1}{2}\sqrt{16p(p-1)+3}  \right)&  p\in\left[3/4,(2+\sqrt{2})/4\right] \\
			1 & p \in\left[(2+\sqrt{2})/4,1\right]\;,
			\end{cases}\notag\\
	&f_{\min}\left(p_t,p\right) = \begin{cases}
	g\left(p\right)&  p \leq p_t \;  \\
	\frac{\mathrm{d}}{\mathrm{d}p} g(p)\big|_{p_t} p+ \Big( g(p_t) -\frac{\mathrm{d}}{\mathrm{d}p} g(p)\big|_{p_t} p_t\Big)& p> p_t\;,
	\end{cases} \nonumber \\
	& v(p_t,\epsilon, \kappa)=2\left( \log_2 9 + \frac{\mathrm{d}}{\mathrm{d}p} g(p)\big|_{p_t}\right)\sqrt{1-2 \log_2 (\epsilon\kappa)}\;, \notag \\
	&n_{\mathrm{EAT},\sigma}(p_t)=\Bigg(\frac{ v(p_t,\epsilon_{\sigma}, \kappa)+\sqrt{ v(p_t,\epsilon_{\sigma}, \kappa)^2+4\sigma f_{\min}\left(p_t,\hat{I}/8+1/2\right)}}{2f_{\min}\left(p_t,\hat{I}/8+1/2\right)}\Bigg)^2\;,\notag
\end{align}
where $h(x)=-x\log_{2}(x)-(1-x)\log_{2}(1-x)$ is the binary entropy function and 
$f_{\min}\left(p_t,p\right)$ with the free parameter $p_t$ is a CHSH-based min-tradeoff function.

We estimate the minimum number of trials required by entropy accumulation with CHSH-based min-tradeoff functions  
when $\sigma=1089$ and $\epsilon_{\sigma}=0.8\times 2^{-64}\approx 4.34\times 10^{-20}$. We observe that the 
smaller the value of $\kappa$, the larger the value of $n_{\mathrm{EAT},\sigma}$ becomes when other parameters 
are fixed. We therefore formally set $\kappa=1$ in the above expression of $n_{\mathrm{EAT},\sigma}$. 
From the first $21\SI{min}$ unblinded data for testing our protocol 
we estimate the expected CHSH violation $(\hat{I}-2)=1.142\times 10^{-3}$. Then $n_{\mathrm{EAT},\sigma=1089}=6.108\times 10^{10}$, which would have taken $169.7\SI{h}$ of experiment time with the trial rate of $100\SI{kHz}$ (this is slightly higher than the trial rate used in the current work). For the DIQRNG implemented with a loophole-free Bell test of Ref.~\cite{liu_yang:qc2018a}, from Tab.~VI therein we estimate the expected CHSH violation $(\hat{I}-2)=2.141\times 10^{-3}$. So, $n_{\mathrm{EAT},\sigma=1089}=1.737\times 10^{10}$, which would have taken $24.1\SI{h}$ of experiment time with the trial rate of $200\SI{kHz}$ used in Ref.~\cite{liu_yang:qc2018a}.



\end{document}